\documentclass[a4paper,11pt]{article}
\pdfoutput=1 

\usepackage{jinstpub} 

\title{\boldmath Calibration of MPD Electromagnetic Calorimeter with Muons}

\author[a,1]{A.Yu.~Semenov,\note{Corresponding author.}}
\author[a]{I.A.~Semenova,}
\author[a,b]{M.~Bhattacharjee,}
\author[c]{A.~Durum,}
\author[a]{Yu.~Krechetov,}
\author[d]{V.~Kulikov,}
\author[e]{I.~Mamonov,}
\author[d]{M.~Martemianov}

\affiliation[a]{Joint Institute for Nuclear Research, 141980, Dubna, Russia}
\affiliation[b]{Gauhati University, Guwahati, Assam 781014, India}
\affiliation[c]{NRC "Kurchatov Institute" - IHEP, 142281, Protvino, Russia} 
\affiliation[d]{NRC "Kurchatov institute" - ITEP, 117218 Moscow, Russia}
\affiliation[e]{P.N. Lebedev Physical Institute, Moscow, 119991 Russia}

\emailAdd{semenov\underline{~}andrei@yahoo.com}

\abstract{Shashlyk-type electromagnetic calorimeter (ECal) of the Multi-Purpose Detector at heavy-ion NICA collider is optimized to provide precise spatial and energy measurements for photons and electrons in the energy range from about 40 MeV to 2-3 GeV. To deal with high multiplicity of secondary particles from Au-Au reactions, ECal has a fine segmentation and consists of 38,400 cells ("towers"). Given the big number of "towers" and the time constraint, it is not possible to calibrate every ECal "tower" with beam. In this paper, we describe the strategy of the first-order calibration of ECal with cosmic muons.}


\proceeding{Calorimetry for High Energy Frontier Conference (CHEF 2019)\\
  November 25-29, 2019\\
  Fukuoka, Japan}

\begin{document}
\maketitle
\flushbottom

\section{Introduction}
\label{sec:intro}
\begin{figure}[htbp]
\centering 
\qquad
\hspace*{0mm}\includegraphics[width=.8\textwidth,origin=a]{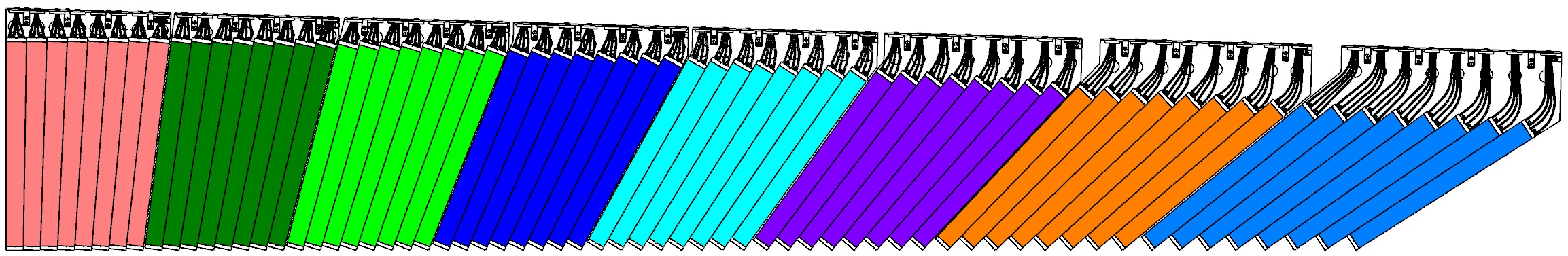}
\caption{\label{project} An evolution of the ECal modules shapes as a function of Z-coordinate (beam direction) in the calorimeter barrel: from module \#0 in the center of ECal (shown in pink in the left side of the figure) to the module \#7 in the edge of ECal (shown in medium-blue in the right side of the figure).}
\end{figure}

The main goal of the Nuclotron-based Ion Collider fAcility (NICA) program at Joint Institute for Nuclear Research (Dubna) is an experimental study of the properties of nuclear matter in the energy region of the maximum baryonic density. One of the NICA detectors, the Multi-Purpose Detector (MPD), is optimized for the study of properties of a hot and dense matter in heavy-ion collisions at $\sqrt{s_{NN}}$=4-11 GeV \cite{mpd}. Electromagnetic barrel calorimeter of MPD (ECal) is designed to provide precise energy and position measurements for photons and electrons in conditions of high multiplicity of secondary particles from the beams interactions. To meet this goal, ECal has a fine segmentation and a projective geometry where each calorimeter cell is oriented to the beams interaction point. In this geometry, exact size of each cell and its position relative to neighbour cells is defined by the cell Z-position (beam direction) in the ECal barrel (see Fig.~\ref{project}). In total, ECal consists of 38,400 cells of 64 different types that are combined into 2,400 modules of 8 different types (shown in Fig.~\ref{project} in different colors).  

\section{Absolute Energy Calibration Method}

The traditional "first-level" absolute calibration of calorimeter contains usually 2 steps: 1) equalize the gain of all calorimeter cells (intercalibration) with usage of electron beam, cosmic muons or physics signals, and 2) find one or few absolute-energy calibration coefficients (for whole calorimeter) from electron beam tests or reconstruction of narrow invariant-mass distributions \cite{cms,g2}. The problem of this method is that it does not take into account geometrical difference of the calorimeter cells and dependency of the calorimeter properties on the energy of the primary particle (viz., works well only in the limited energy range) and the locations of the cell and the particle hit. Attempts to correct on these effects result in additional labour-consuming measurements and "phenomenological" calibrations on impact particle energy, p$_T$, hit-position, and other non-linear dependencies \cite{jones}. Occurrence of the energy dependence is due to the fact that the composition of the electromagnetic shower changes as a function of depth, or age. In the late stages, most of the energy is deposited by soft gammas which undergo Compton scattering or photoelectric absorption, and the sampling fraction for this shower component (i.e., the fraction of the energy that contributes to the calorimeter signals) is visibly less than that of the MIPs that dominate the early stages of the shower development \cite{wigmans}. MPD ECal has limited thickness of about 11X$_0$, so increase of the primary particle energy results in the moving of the maximal-sampling-fraction region closer to the photodetector as well as moving the late-stage shower development region out of the calorimeter; that effect also minimize the absorbed-in-fibers fraction of the produced scintillation light. The reason for the dependence on the Z-position of the primary particle hit is that the calorimeter with projective geometry consists of 64 types of cells with different shapes and different surrounding of the neighbour cells, so the different fraction of the shower from different stages of development goes into neightbour cells and leaks out of the calorimeter. 

\begin{figure}[htbp]
\centering 
\qquad
\hspace*{0mm}\includegraphics[width=.6\textwidth,origin=a]{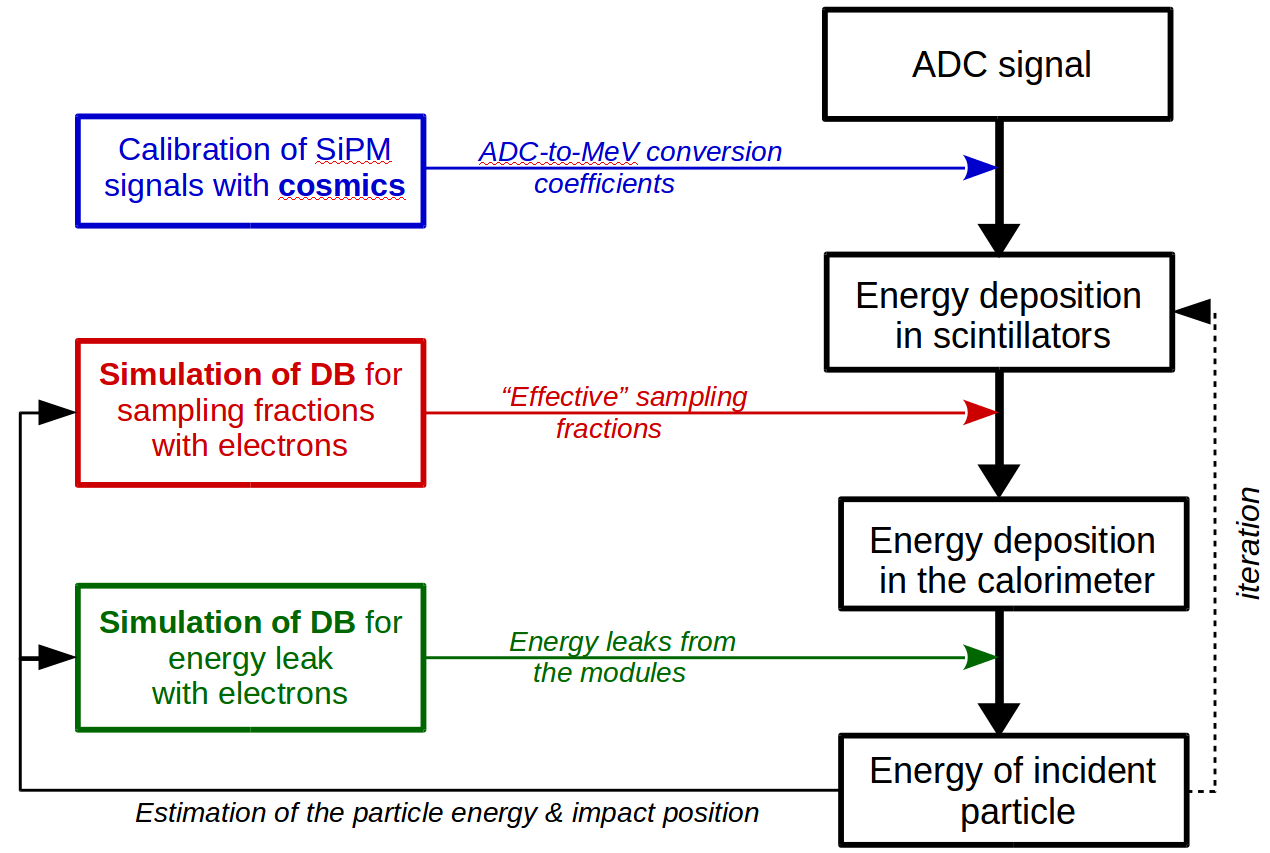}
\caption{\label{flow} The flow diagram of the iterative event-by-event calibration of ECal data.}
\end{figure}
Our calibration approach (see Fig.~\ref{flow}) consists of two steps also. First, we plan to establish connection between the amplitudes of  observed signals and the "effective" energy depositions in the active volume of the calorimeter (scintillators) for each calorimeter cell; it can be done via comparison of the calorimeter signals from cosmic muons with predictions from corresponding detailed simulation. The second step contains computer simulations to create a data base (DB) of the physical properties of the calorimeter (viz., sampling fractions, energy leaks etc.) as a functions of primary particle energy, hit position, location of the cell in the calorimeter etc. to convert the energy deposition in the calorimeter active volume to the impact particle energy, and do it in phenomenological-dependent-free manner. Another big advantage of the proposed method is that no beam is needed for the absolute energy calibration of the calorimeter. With detailed description of the ECal structure in the simulation (viz., exact structure of ECal towers, shapes, locations as well as measured light absorption in the WLS fibers), we hope to perform the preliminary ECal calibration with accuracy of about 3\% or better.  

\section{Computer Simulations}

\begin{figure}[htbp]
\centering 
\qquad
\hspace*{0mm}\includegraphics[width=.47\textwidth,origin=a]{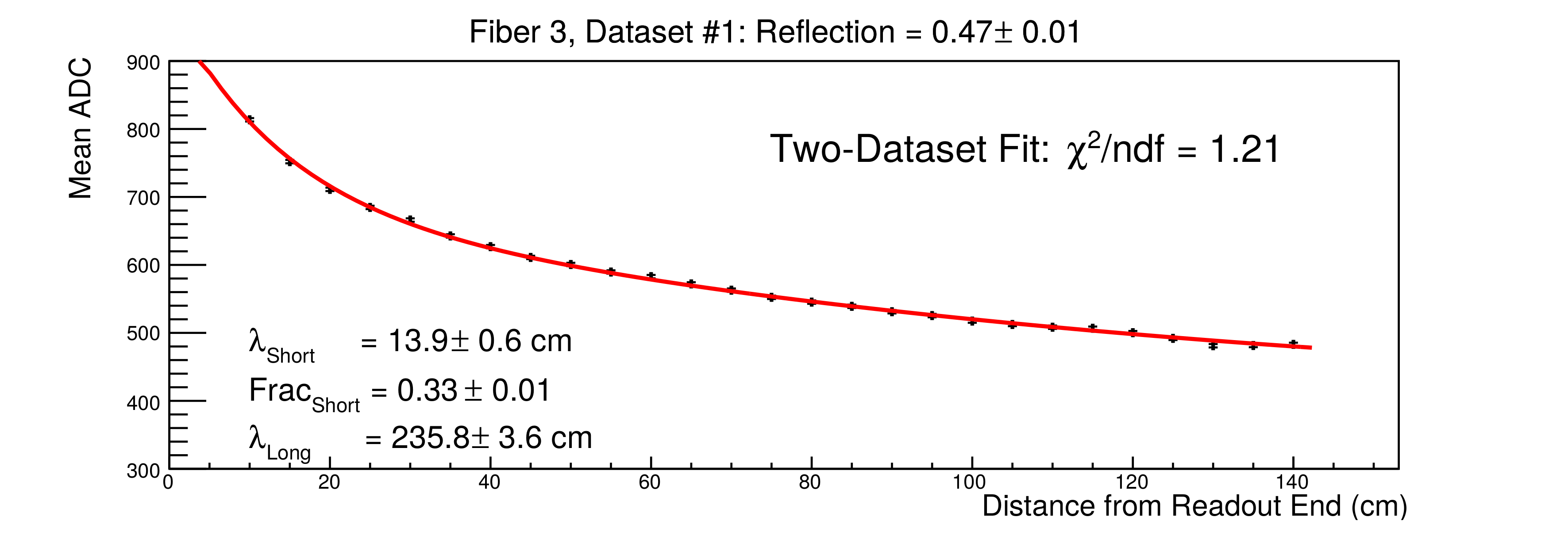}
\hspace*{0mm}\includegraphics[width=.47\textwidth,origin=b]{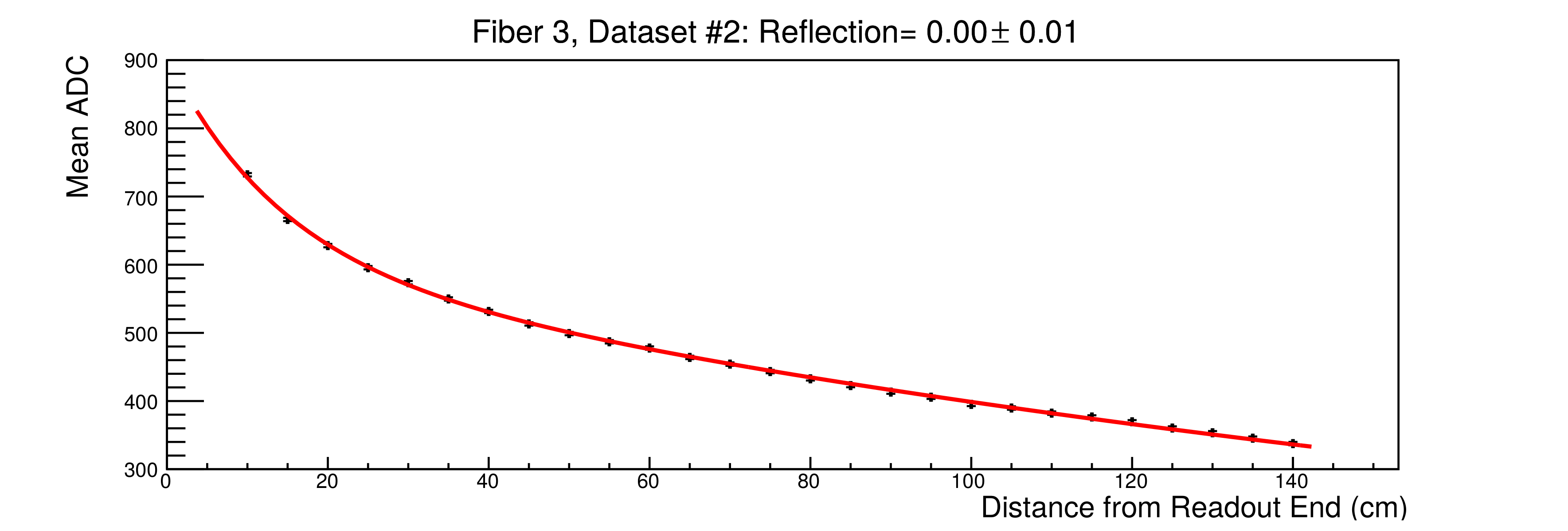}
\caption{\label{att} Left panel: attenuation of the scintillation light in the 150-cm WLS fiber with white reflective paint on the "far" end. Right panel: attenuation of the light in the same fiber with no reflection from the "far" end. Simultaneous fit of both data sets with correspondent two-attenuation-lengths functions. Details of the measurements can be found in the paper \cite{durum}.}
\end{figure}

Both steps of the calibration require detailed computer models of the calorimeter modules. 
The calorimeter has the projective geometry, so the shape of each cell and it's location relative to the neighbor cells depend on the Z-coordinate (along the beam) of the cell in the calorimeter barrel.
Each module contains 2$\times$8=16 calorimetric cells, and there are 8 module types of different shapes. Each calorimeter cell consists of 210 layers of 1.5-mm-thick plastic scintillator interlaid with 0.3-mm-thick lead plates. 16 WLS fibers are passed through the holes in the lead and scintillators to collect the light and transport it to the photodetector. The "far" end (from the photodetector) of each fiber as well as lead plates and the sides of the scintillators are painted with white reflective paint to maximize the light collection. The attenuation of the light in the WLS fibers was carefully measured using the technique described in \cite{durum} (see Fig.~\ref{att}). 
Taking into account the significant attenuation of the scintillation light along the WLS fibers, a convolution of the energy deposited into the scintillators with the attenuation function (viz., an "effective" energy deposition that is proportional to the light transported to the photodetector) represents better the calorimeter signal than just energy deposited into scintillators, so the former was used for estimation of "effective" sampling fractions etc. in the simulations.

The simulation was done with FLUKA version 2011.2x.7 Monte Carlo code \cite{fluka} with energy cuts and transport parameters according to one of the default FLUKA set CALORIMETRY (with minor tune).

\section{Measurements with Cosmic Muons}

\begin{figure}[htbp]
\centering 
\qquad
\hspace*{-10mm}\includegraphics[height=.3\textwidth,origin=a]{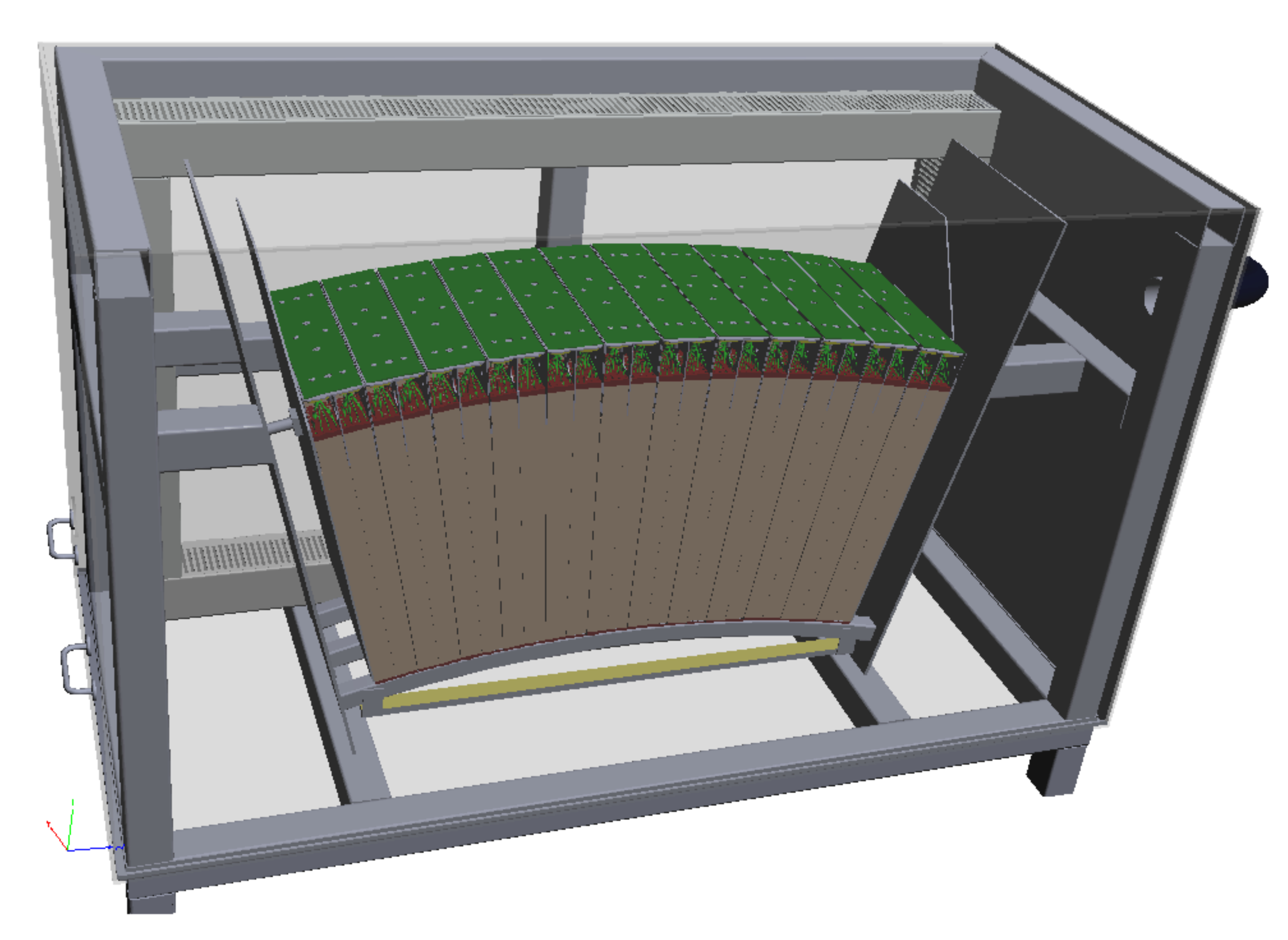}
\hspace*{12mm}\includegraphics[height=.25\textwidth,origin=b]{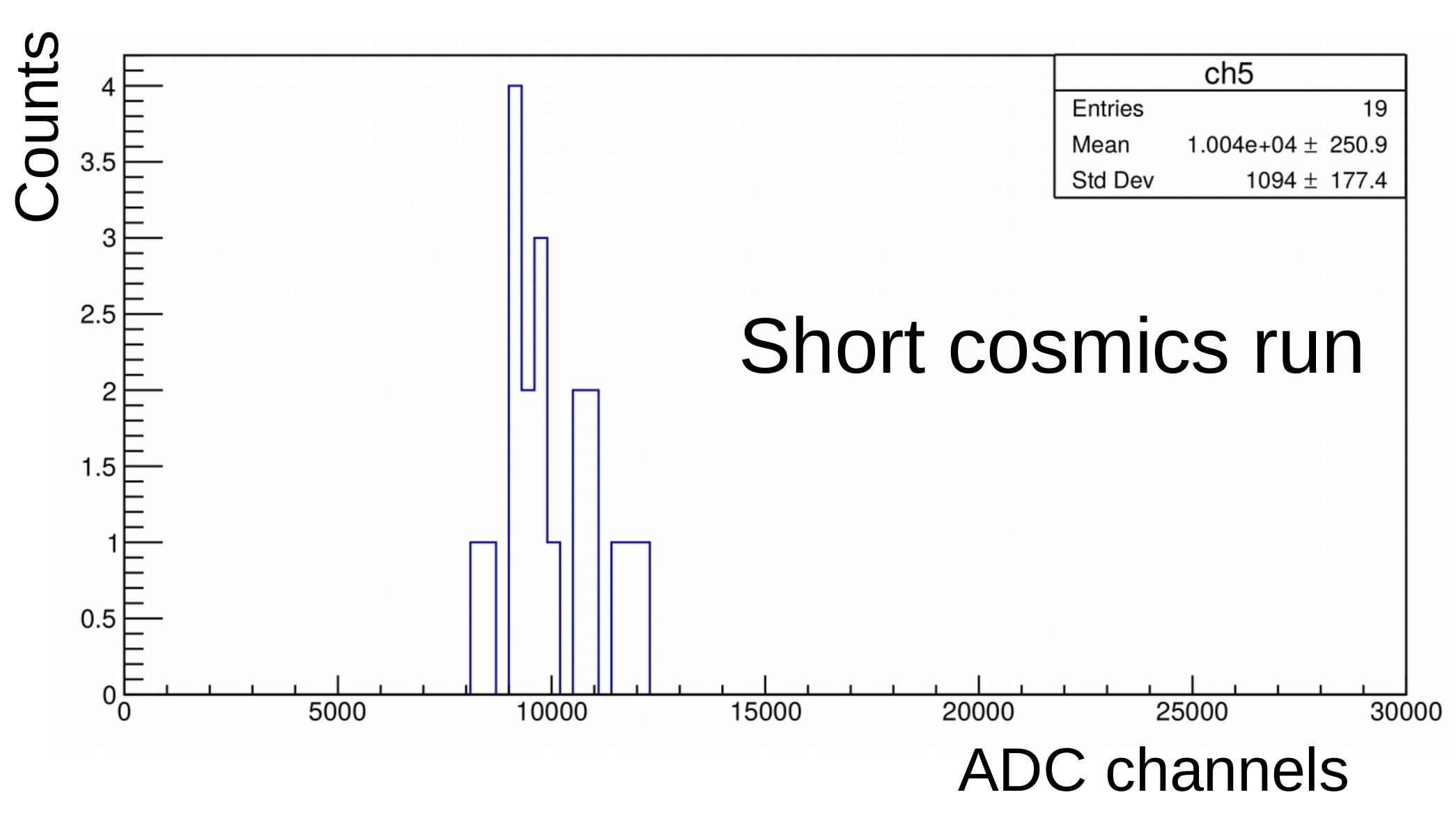}
\caption{\label{stand} Left panel: the stand to test simultaneously 12 ECal modules. Right panel: the compact spectrum of signals from cosmic muons from a short (a few hours) run.}
\end{figure}

For the measurements with cosmic muons, we orient the modules vertically.
To select the "longitudinal" flux of muons along each cell axis (similar to that is expected during the calorimeter work on the beam), 
we use external scintillation trigger detectors as well as "veto" signals from neighbour cells. Compared to usage of "transverse" muons, the dispersion of the spectrum measured with "longitudinal" orientation of the modules is insensitive to the cell's trapezoidal shape and visible light absorption in it; that results into more compact (in compare with other module orientations) distributions (see right panel of Fig.~\ref{stand}). The downside here is a relatively low rate of selected muons; though, the muon statictics collected in 2 weeks is expected to be enough for the measurement of ADC-to-energy coefficients with an accuracy better than 1\%.

To perform the measurements with cosmic muons, the special stands were developed (left panel of Fig.~\ref{stand}). Each stand allows simultaneous measurements of 12 modules; so 8 stands (for 8 different types of modules) allow the calibration of 2,400 ECal modules in one year. We plan to combine the calibration measurements with the produced modules quality assurance (QA). 

The measured with cosmics spectra will be compared with the simulated spectra for the energy that is deposited by muons in the scintillator plates in the modules and corrected on the attenuation of the scintillation light on it's way to the photodetectors (viz., the energy that is "visible" in the calorimeter), and the ADC-to-"visible" energe coefficients will be obtained for each calorimeter cell.

\section{Sampling Fraction and Energy Leak}

\begin{figure}[htbp]
\centering 
\qquad
\hspace*{-5mm}\includegraphics[width=.43\textwidth,origin=a]{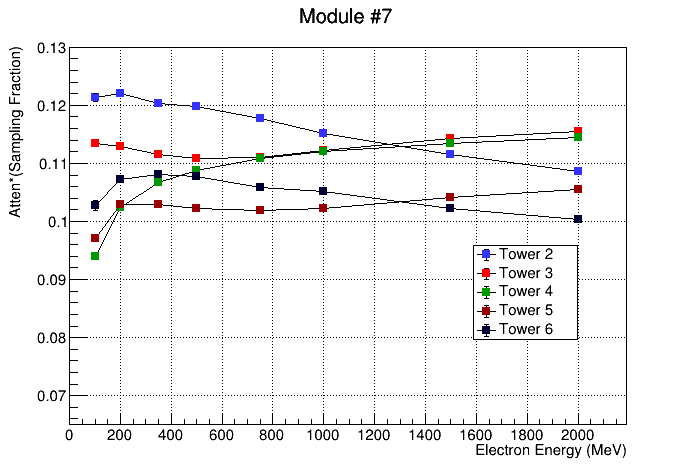}
\hspace*{5mm}\includegraphics[width=.43\textwidth,origin=b]{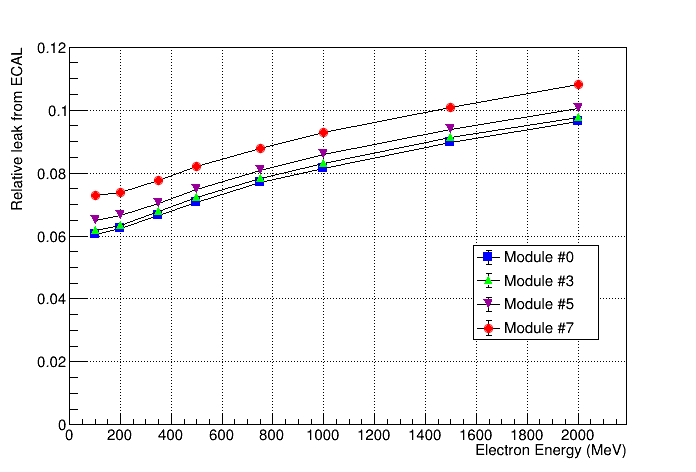}
\caption{\label{sf} Left panel: the "effective" (viz., convoluted with attenuation function) sampling fraction for ECal module of type \#7; the electron beam impacts the middle of the cell (tower) \#4. Right panel: relative energy leak for electron beam impact in the center of different ECal modules.}
\end{figure}
The pilot computer simulations of electromagnetic showers in the ECal modules were performed. Fig.~\ref{sf} shows "effective" sampling fraction (viz., the sampling fraction that is convoluted with light attenuation function) and energy leak from the calorimeter, and indeed these vital ECal parameters depend visibly on the impact particle energy, module/cell type, and distance from the cell of interest to the particle impact point as expected.

\section{Plans and Conclusions}

We plan to perform ECal calibration procedure that relies on "energy scale" measurements with MIPs (muons) and detailed simulation of
the calorimeter parameters. Production of the data base calibration parameters requires massive
simulations in 2020 and 2021. The simulated calibration parameters are the functions of the incident particle
energy, cell location and impact position, so no additional non-linear corrections are required. We plan to perform calibration on muons for all ECal modules during 2020 and 2021 (together with the modules QA). Preliminary test of the calibration procedure was done with 50-500~MeV electron beam at  Lebedev Physics Institute of Russian Academy of Science (Troitsk, Moscow reg., Russia); the data analysis is in progress.

\acknowledgments

This work was supported by the Russian Foundation for Basic Research (RFBR) according to the research projects No~18-02-40079 and No~18-02-40083.


\end{document}